\documentclass[english,aps,prl,twocolumn,showpacs,superscriptaddress,groupedaddress,footinbib]{revtex4}  
\usepackage{graphicx}  
\usepackage{dcolumn}   
\usepackage{bm}        
\usepackage{amssymb}   
\usepackage{babel}
\usepackage{verbatim}
\usepackage{amsmath}
\usepackage{setspace}

\graphicspath{{figures/}}

\usepackage{siunitx}

\begin{document}

\widetext

\title{Spin ensemble-based AC magnetometry using concatenated dynamical decoupling at low temperatures}

\author{D. Farfurnik}
\affiliation{Racah Institute of Physics, The Center for Nanoscience and Nanotechnology, The Hebrew University of Jerusalem, Jerusalem 9190401, Israel}

\author{A. Jarmola}
\affiliation{Department of Physics, University of California, Berkeley, California 94720-7300, USA}

\author{D. Budker}
\affiliation{Helmholtz Institute, JGU, Mainz, Germany}
\affiliation{Department of Physics, University of California, Berkeley, California 94720-7300, USA}

\author{N. Bar-Gill}
\affiliation{Racah Institute of Physics, The Center for Nanoscience and Nanotechnology, The Hebrew University of Jerusalem, Jerusalem 9190401, Israel}
\affiliation{Dept. of Applied Physics, Rachel and Selim School of Engineering, Hebrew University, Jerusalem 9190401, Israel}

\date{\today}

\begin{abstract}
	Ensembles of nitrogen-vacancy (NV) centers in diamond are widely used as AC magnetometers. While such measurements are usually performed using standard (XY) dynamical decoupling (DD) protocols at room temperature, we study the sensitivities achieved by utilizing various DD protocols, for measuring magnetic AC fields at frequencies in the 10-250 kHz range, at room temperature and 77 K.  By performing measurements on an isotopically pure $^{12}$C sample, we find that the Carr-Purcell-Meiboom-Gill (CPMG) protocol, which is not robust against pulse imperfections, is less efficient for magnetometry than robust XY-based sequences. The concatenation of a standard XY-based protocol may enhance the sensitivities only for measuring high-frequency fields, for which many ($> 500$) DD pulses are necessary and the robustness against pulse imperfections is critical. Moreover, we show that cooling is effective only for measuring low-frequency fields ($\sim 10$ \SI{}{\kilo\hertz}), for which the experiment time apporaches $T_1$ at a small number of applied DD pulses. 
\end{abstract}
\maketitle

\paragraph{}
The unique spin and optical properties of the negatively-charged nitrogen-vacancy (NV) color centers in diamond place them as leading platforms for quantum-information processing \cite{Cappellaro2009,Bennett2013,Weimer2013} and sensing \cite{Taylor2008, Balasubramanian2008, Pham2012, Mamin2014,Dolde2011}. Since the amount of fluorescent light emitted from these defects is proportional to the number of spins,  ensembles consisting of many NV centers are typically used for efficient magnetometry \cite{Taylor2008, Balasubramanian2008, Pham2012}. When the magnetic field of interest is oscillating at a certain frequency (AC field), enhanced sensitivities can be achieved using dynamical decoupling (DD) microwave (MW) pulse sequences. These DD sequences improve the coherence times of the NV ensemble, and their proper synchronization with the AC field enables phase accumulation of the spin state, resulting in enhanced magnetometric sensitivities. 
However, the efficient time for phase accumulation is limited by the coherence time ($T_2$) of the NV ensemble, and the measured fluorescence contrast drops due to the accumulation of DD pulse errors \cite{Wang2012a,Wang2012,Farfurnik2015}. Therefore, for different given diamond samples and AC frequencies to be measured, different DD protocols and working temperatures may produce optimal magnetometric sensitivities. While, in most cases, standard (XY) sequences are implemented at room temperature for AC magnetometry, here we identify the higher-performance protocols and operation temperatures for measuring a broad-frequency range of AC fields, using an ensemble of $\sim 10^4$ NV centers. We experimentally demonstrate the sensitivity enhancement, within a range of AC frequencies between $10$ kHz and $250$ \SI{}{\kilo\hertz}.

\begin{figure}[]	
	\includegraphics[width=1\columnwidth]{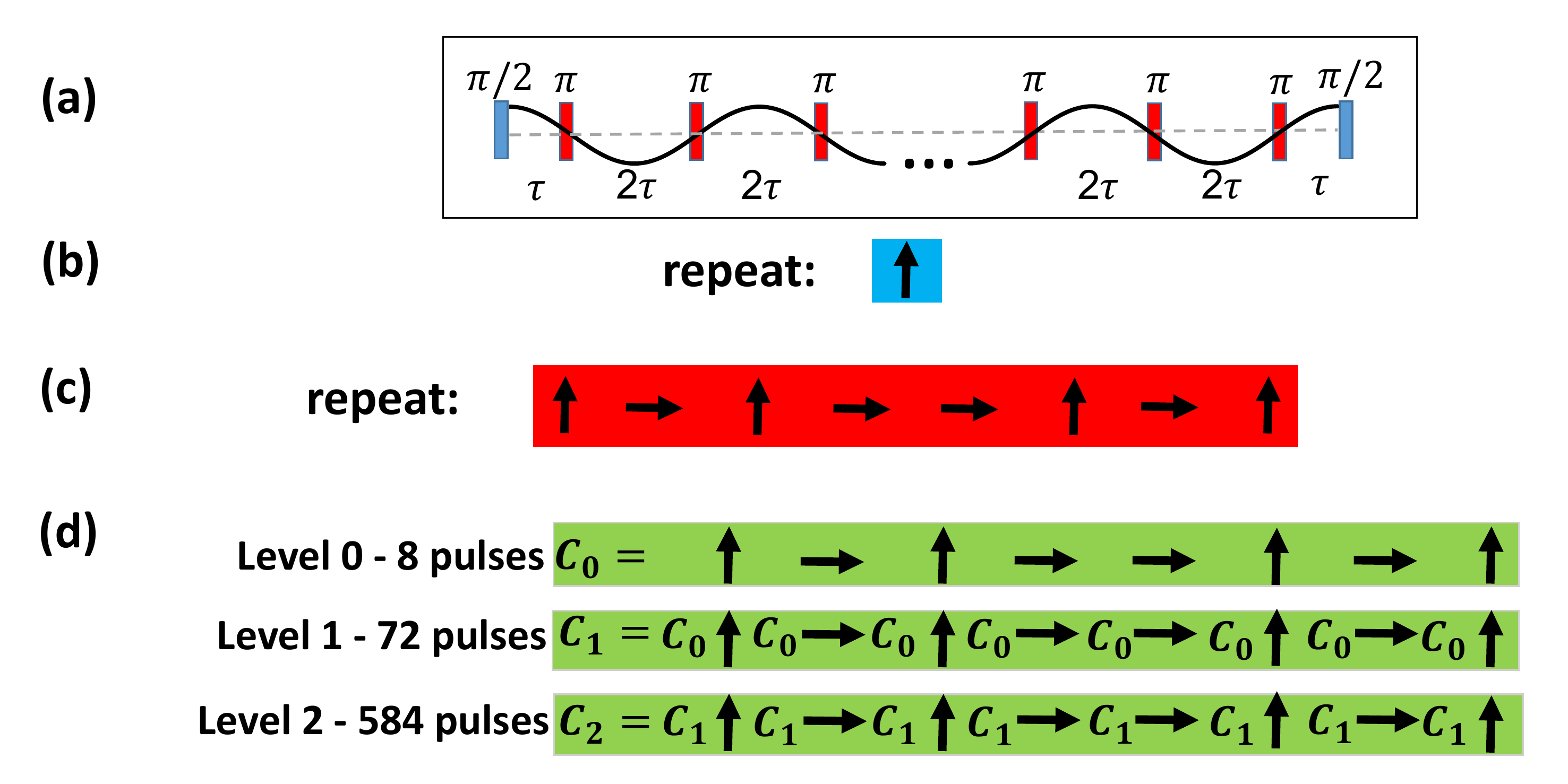}
	\caption{(Color online) (a) AC Magnetometry scheme: DD $\pi$-pulses are synchronized with the frequency of the fluctuating field. The state is initialized along the ``X" axis, and read out at the end of the sequence by the flourescence emitted after applying a 532-nm laser pulse. For the two conventional pulse sequences, (b) CPMG - all pulses are applied along the same axis, and (c) XY8 - pulses are applied at two perpendicular (``X" and ``Y") phases, the basic structure is repeated.  In the last sequence (d), the concatenated version of XY8, the phases of the pulses are determined recursively from previous experiments. The directions of the arrows represent the phases of the pulses.} 
	\label{fig:DDschemes}
\end{figure}
\paragraph{}
The coherence time of NV centers is limited by two main factors: The longitudinal relaxation ($T_1$), which is typically dominated at room temperature by phononic interactions, and can be reduced by three orders of magnitude using modest cooling to about 80 K; \cite{Jarmola2012}, and the "spin-bath" environment, creating randomly fluctuating magnetic fields, which is typically dominated by
$^{13}$C nuclear spins and nitrogen paramagnetic spin impurities \cite{deSousa2009,BarGill2012}. DD pulse sequences [Fig. \ref{fig:DDschemes}(a)]
can be used to suppress the effect of these fluctuations and thus lead to long NV spin coherence times \cite{BarGill2012,Bargill2013}. 
Since experimental conditions are not ideal, off-resonant driving due to the NV hyperfine structure, as well as pulse imperfections, may cause a rapid drop in the signal contrast with the number of applied DD pulses, significantly reducing the ability to preserve arbitrary spin states \cite{Wang2012a,Wang2012,Farfurnik2015}. Recently, the robustness of various DD protocols against such imperfections was examined. By applying a simple concatenation process \cite{Khodjasteh2005,Witzel2007} on standard DD sequence (such as XY8), an arbitrary spin state of an NV ensemble was preserved up to $T_2 \sim 30$ ms \cite{Farfurnik2015}, longer than achieved by conventional Carr-Purcell-Meiboom-Gill (CPMG) [Fig. \ref{fig:DDschemes}(b)] \cite{Meiboom1958} and XY-based [(Fig. \ref{fig:DDschemes}(c)] \cite{Gullion1990} DD sequences, which are less robust to these imperfections. In contrast to the conventional DD protocols, in which a single structure of the phases of the $\pi$-pulses is repeated many times, the main idea behind the concatenated protocols is that the phases of the pulses are built recursively from previous experiments (``levels of concatenation")  [Fig. \ref{fig:DDschemes}(d)] \cite{Khodjasteh2005,Witzel2007}.

\begin{figure}[!t]	
 	\includegraphics[width=1\columnwidth]{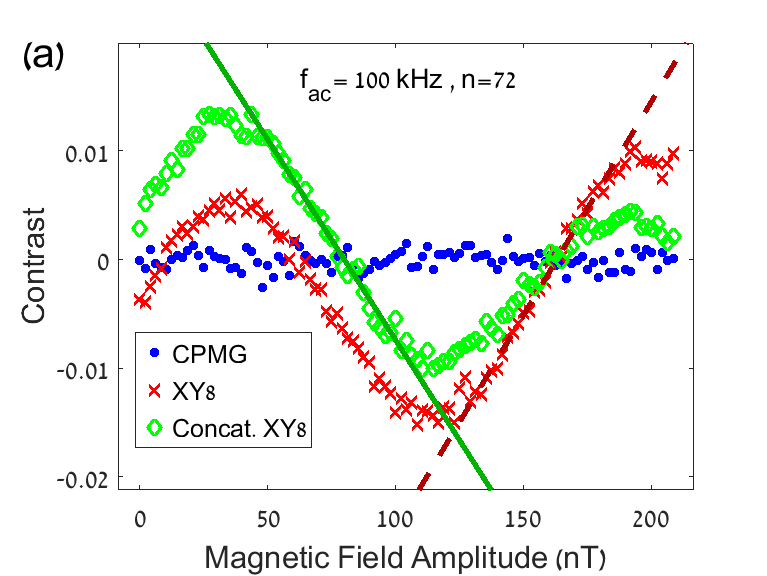} 
 	\includegraphics[width=1\columnwidth]{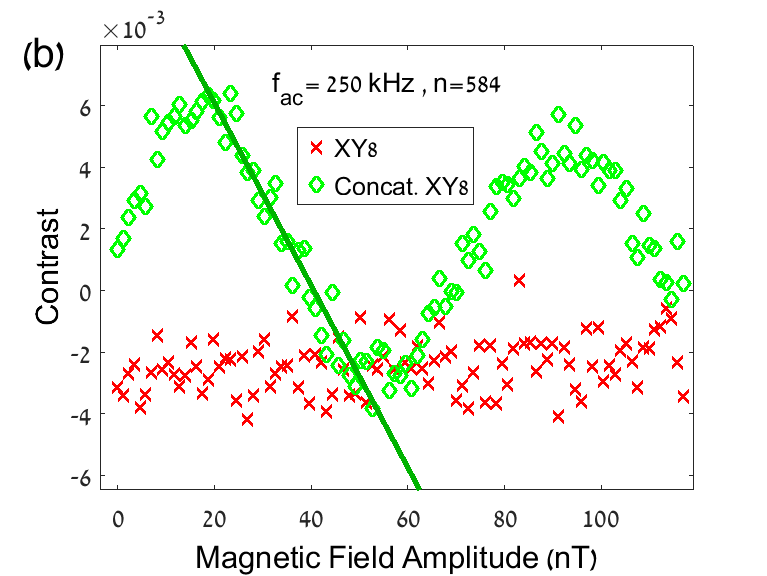} 
 	\caption{(Color online) Measured fluorescence contrast as a function of the AC field amplitude, utilizing the CPMG, XY8 and concatenated XY8 DD protocol at room temperature. The frequency of the fluctuating field and the number of applied pulses are (a) $f_{AC}=100 $ \SI{}{\kilo\hertz}, $n=72$ and (b)  $f_{AC}=250 $ \SI{}{\kilo\hertz}, $n=584$. The lines represent the largest slopes, which determine the magnetic field amplitude sensitivities. The results of different frequencies and numbers of pulses are given in the supplemental material \cite{Suppl}.}
 	\label{fig:DDresults}
 \end{figure}

\begin{figure}[!t]	
	\includegraphics[width=1\columnwidth]{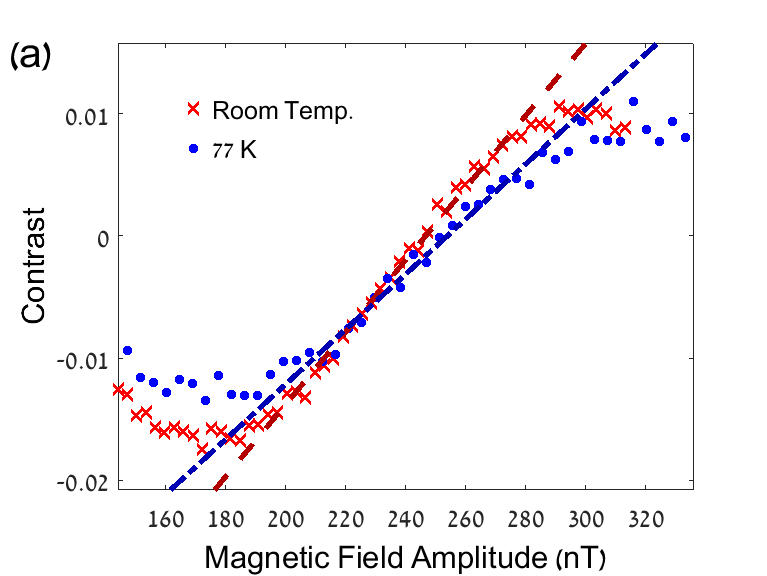} 
	\includegraphics[width=1\columnwidth]{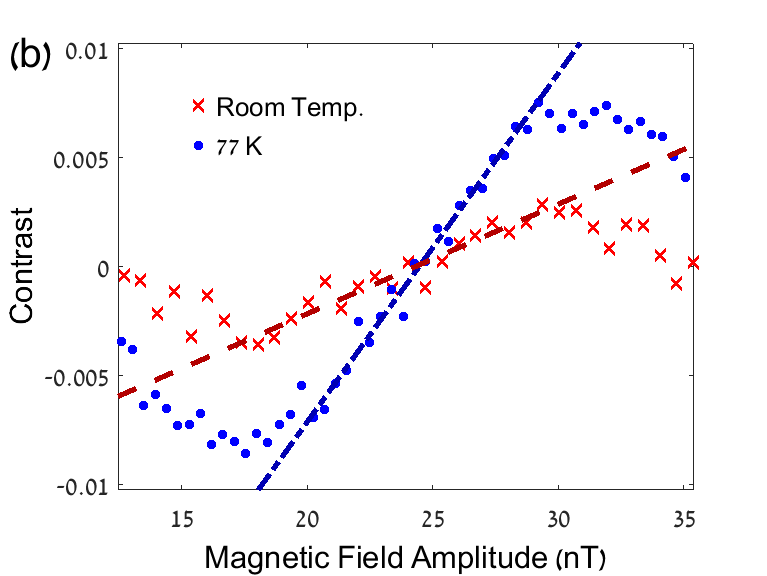} 
	\caption{(Color online) Measured fluorescence contrast as a function of the AC field amplitude, utilizing the XY8 DD sequence with $n=48$ pulses at room temperature and $77$K. The frequencies of the fluctuating fields are (a) $f_{AC}=100 $ \SI{}{\kilo\hertz} and  (b)  $f_{AC}=10 $\SI{}{\kilo\hertz}. The lines represent the largest slopes, which determine the amplitude magnetic field amplitude sensitivities.}
	\label{fig:Tempresults}
\end{figure}

\paragraph{}
Let us assume an AC magnetic field fluctuating with a known frequency $f_{AC}$, and an unknown amplitude $B_{AC}$ to be measured:
\begin{equation}
B=B_{AC}\cdot \sin\left(2\pi f_{AC}\cdot t + \phi\right).
\end{equation}
If the spin state of the NV ensemble is initialized along the $X-Y$ plane of the Bloch sphere using a resonant MW $\frac{\pi}{2}$-pulse, the spin will oscillate according to the field intensity and frequency. The accumulating phase of the precessing spin will be directly proportional to the strength of the field. As a result, the strength of the field can be derived after applying an additional $\frac{\pi}{2}$-pulse, followed by a fluorescence measurement. The magnetometric sensitivity achieved from such a ``Ramsey" cycle is poor: it is limited by the natural dephasing time $T_2^*$ \cite{Taylor2008}. Moreover, after half an oscillation period, the spin will precess in the opposite direction, leading to the cancellation of the accumulated phase and reduction in the sensitivity. In order to overcome these obstacles,  DD sequences can be applied  [Fig. \ref{fig:DDschemes}(a)]: as mentioned previously, DD sequences increase the coherence time $T_2$. In addition, the application of  $\pi$-pulses in synchronization with the fluctuating field (every half-period of the field) enables phase accumulation within the whole coherent time interval, significantly enhancing the magnetometric sensitivities.   
Under these conditions, the theoretical magnetometric sensitivity using DD is given by \cite{Pham2012}:
\begin{equation}
\eta(n)=
\frac{\pi\hbar}{2g\mu_B}\frac{1}{C^{(n)}\sqrt{\frac{n}{2f_{AC}}}}exp\left[\left(\frac{n^{(1-s)}}{2T_2^{(1)}f_{AC}}\right)^p\right],
 \label{equation:theosens}
\end{equation}
where $T_2^{(1)}$ is the coherence time for a single (``Hahn-Echo") $\pi$-pulse, $n$ is the number of DD pulses, $s$ is the scaling factor of the coherence time with the number of pulses, ($T_2^{(n)}=T_2^{(1)}n^s$, until approaching saturation limited by $T_1$), $p$ is a stretching factor of the decay, which can be derived from the functional form of the experimentally measured decoherence curve ($\propto \exp\left[({-t/{T_2^{(n)}})^p}\right]$), and $C^{(n)}$ represents the optical collection efficiency, the signal-to-noise ratio directly proportional to the square-root of number of NVs, and the measured fluorescence contrast. Since the experiment time grows linearly with the number of pulses, the coherence time is eventually limited by $T_1$, and assuming an ideal experimental apparatus, one can derive an analytical expression for the optimal number of pulses $n$ for any frequency $f_{AC}$ \cite{Pham2012}. However, due to the accumulation of pulse errors, the signal contrast may reduce with the number of pulses in a nontrivial way \cite{Wang2012,Wang2012a,Farfurnik2015}, which may have a significant effect on the achieved sensitivities. 
\paragraph{}
In order to experimantelly measure the unknown amplitude of the AC field using a diamond sample with a given NV ensemble, DD $\pi$-pulses can be synchronized with the field. The sensitivity of such a measurement can be obtained by continuously monitoring the amplitude of the field, resulting in different phase accumulations and a fluorescence signal oscillating with the field's amplitude. The resulting magnetometric sensitivity can be obtained by \cite{Pham2012}
 \begin{equation}
 \eta(n)=\frac{\sigma}{(\delta S / \delta B)}\sqrt{T},
 \label{equation:magsens}
 \end{equation}
 where $T$ is the total experiment time, $\sigma$ is the standard error in the measured signal, and $(\delta S / \delta B)$ is the maximal change in the signal caused by a small deviation in the amplitude of the AC field, which can be derived from the largest slope in the oscillating signal. In this work, such an analysis is performed for AC fields with various frequencies ranging from 10 kHz to 250 kHz, and various DD protocols are examined at room temperature and 77 K. 
\paragraph{}
Our measurements were performed on a $^{12}$C enriched ($99.99\%$) diamond sample, grown via chemical vapor deposition, having nitrogen concentration of $\sim 2 \times 10^{17}$ $\SI{}{\centi\meter}^{-3}$ and NV concentration of
$\sim 4 \times 10^{14}$ $\SI{}{\centi\meter}^{-3}$ (Element Six). A 532-nm laser was used to optically-excite an ensemble consisting of $10^4$ NV centers within a $\sim 25$  $\SI{}{\micro\meter}^{3}$
measurement volume. A permanent magnet producing a static magnetic field of $B_0 \sim 300$ G along the NV symmetry axis was used to Zeeman-split the $m_s=\pm 1$ spin sublevels. A $70-\SI{}{\micro\meter}$-diameter wire was used to manipulate the ensemble's spin state with the $m_s=0\rightleftarrows m_s=+1$ transition. The spin-rotation axes of the applied DD pulses
were set through in-phase and quadrature (I/Q) modulation of the MW carrier signal from the signal generator (SRS SG384).
The AC field of interest was introduced by an additional signal generator (SRS SG345), which produced AC currents into a home-made coil through resistors of $50$ or $1400$ $\Omega$, which were chosen considering the desired frequencies and amplitudes of the AC field. At room temperatures, the measured Hahn-Echo coherence time was $T_2 \sim 270$ $\SI{}{\micro\second}$ and the longitudinal relaxation time was  $T_1 \sim 5$ $\SI{}{\milli\second}$. For the cryogenic measurements, we used a continuous-flow cryostat (Janis ST-500) and cooled down to 77 K using liquid nitrogen. 
\paragraph{}
The amplitudes of the AC field at the sample position [horizontal axes in Figs. \ref{fig:DDresults},\ref{fig:Tempresults}] were obtained by calibrating the frequencies of the oscillating fluorescence signal: for a known frequency $f_{AC}$ and an unknown amplitude $B_{AC}$, the accumulated phase of the ensemble spin state during a single period of the field is $\phi=4\gamma B_{AC}/f_{AC}$, where $\gamma$ is the NV electronic-spin gyromagnetic ratio \cite{Pham2012}. As a result, a measured oscillation period of $\phi=2\pi$ corresponds to an amplitude difference of $\Delta B_{AC} =(\pi f_{AC})/(2\gamma)$, which can be used to calibrate the AC field amplitudes. These amplitudes cannot be easily estimated by theoretically calculating the resulting magnetic fields given the AC currents: the sample is placed inside a cryostat consisting of metal parts, resulting in different non-trivial attenuations of the magnetic field reaching the samples for different AC frequencies. The fluorescence signal contrast, representing the phase accumulation of the spin-state, was measured by projecting the final states toward the axis perpendicular to the initialization \cite{Suppl}. 
\paragraph{}
The fluorescence signals as a function of the amplitude of the AC field are shown in Fig. \ref{fig:DDresults} for the examined DD protocols and two different AC frequencies and numbers of DD pulses at room temperature. Additional results are given in the supplemental material \cite{Suppl}.
 Since all $\pi$-pulses in the CPMG protocol are applied along the same axis, this protocol cannot preserve arbitrary spin states \cite{Farfurnik2015}. This leads to a dramatic reduction in the signal contrast, which drops below the noise floor even at small numbers of applied pulses [Fig. \ref{fig:DDresults}(a)], making the CPMG protocol inefficient for AC magnetometry. For the XY8 sequence, in which pulses are applied along two perpendicular axes, oscillations in the signal contrast with the magnetic field are seen clearly observable. 
 \begin{center}
 	\begin{table}[t]
 		\small
 		\begin{tabular}{| c | c | c | c | c |}
 			\hline
 			$f_{AC}$ $[\SI{}{\kilo\hertz}]$ &  10 & 25  & 100 & 250   \\ 			
 			\hline
 			$\eta$ [nT$/\sqrt{\SI{}{\hertz}}]$ & $9(1)$ & $10(1)$  & $12(1)$ & $20(2)$   \\ 			
 			\hline								
 			$n$ & 32 & 48  & 144 & 144   \\ 			
 			\hline							
 		\end{tabular}
 		\caption{Highest magnetometric sensitivities, evaluated according to Eq. (\ref{equation:magsens}), utilizing the XY8 sequence and the optimal number of $n$ $\pi$-pulses for each AC frequency. The main sources for errors are $\sim 10\%$ uncertainties in extacting the maximal slopes $(\delta S / \delta B)$.}
 	\end{table}
 \end{center} 
 \paragraph{}
 By calculating the sensitivities from Eq. (\ref{equation:theosens}) and identifying the optimal number of pulses for each frequency (summarized in Table I, full results in the supplemantal material \cite{Suppl}), we conclude that larger numbers of pulses are required for achieving the optimal sensitivities for higher AC frequencies. These results are expected since the applied $\pi$-pulses are synchronized with the period of the AC field based on Eq. (\ref{equation:theosens}), the resulting optimal sensitivities should be better for higher frequencies. Experimentally however, since the XY8 sequence is not fully robust against pulse imperfections, the signal contrast drops with the number of pulses \cite{Suppl}, significantly reducing the sensitivities for higher AC frequencies, and resulting in the opposite effect (better sensitivities for lower frequencies). Moreover, if the optimal number of pulses for measuring a field of 100 kHz is $n=144$, one would expect that a larger number of pulses is required to produce optimal results for measuring a field of 250 kHz. Experimentally, however, due to the accumulation of pulse imperfections, the signal contrast utilizing the XY8 sequence with more than 144 pulses drops below the noise floor [for example, Fig. \ref{fig:DDresults}(b) and full results in the supplemental material \cite{Suppl}], making it no longer effective for magnetometry.
 
 \paragraph{}
 As shown in previous work on spin ensembles, a concatenation procedure on the conventional XY8 sequence results in enhanced robustness against the accumulation of pulse imperfections \cite{Farfurnik2015}. Although such a concatenation process does not exhibit any advantage over the conventional sequence for measuring AC fields at small numbers of pulses [e.g. $n=72$, Fig. \ref{fig:DDresults}(a)], it leads to a significant enhancement in the signal contrast and the resulting magnetometric sensitivities for large numbers of pulses [e.g. $n=584$, Fig. \ref{fig:DDresults}(b)], for which the conventional XY8 sequence is no longer effective. Concatenation is a recursive process, thus the two smallest numbers of pulses applicable in the concatenation of the XY8 protocol is $n=72$ (``first level of concatenation") and $n=584$ (``second level of concatenation") [Fig. \ref{fig:DDschemes}(d)]. As a result, we expect that for AC frequencies higher than examined in this work, for which the number of optimal pulses for a DD protocol is larger than 500, the concatenated version of the DD process will outperform its conventional application. Such an enhanced performance is already seen in Fig. \ref{fig:DDresults} (b), although this is not the optimal number of pulses for the shown frequency. Additionally, changes in the signal contrast due to the accumulation of pulse imperfections may also lead to asymmetric non-trivial phase accumulation during the spin evolution.  In particular, even for very weak AC fields, the application of hundreds of $\pi$-pulses leads to non-ideal refocusing of the initial state (contrast different than 0). This effect is less significant for the concatenated XY8 sequence, which is more robust against pulse imperfections. 
 
\paragraph{}
Finally, we examine the contribution of cooling the diamond sample for the possible enhancement of the magnetometric sensitivities [Fig. \ref{fig:Tempresults}]. At room temperature, the longitudinal relaxation time of $T_1 \sim $ 5 ms limits the achieveable coherence times. The experimentally measured coherence time utilizing the XY8 DD protocol with 48 $\pi$-pulses is $T^{room}_2(48)\approx 1.2$ ms. By  liquid-nitrogen cooling of the sample down to 77 K, the longitudinal relaxation time is extended by more than three orders of magnitude \cite{Jarmola2012}, no longer limiting the coherence times for any practical purpose. As a result, the experimentally measured coherence time utilizing a DD protocol with 48 pulses at 77 K is $T^{77 K}_2(48)\approx 4$ ms. When such a DD sequence is used for measuring an AC field with a frequency of $100$ kHz, the total time of a single experiment is $n/(2f_{AC})\approx 0.24$ ms, much shorter than the coherence times both at room and cold temperatures. In such a case, both measurements at room temperature and 77 K lead to similar maximal slopes of the oscillating signal  [Fig. \ref{fig:Tempresults}(a)], resulting in similar magnetometric sensitivities. However, for measuring an AC field with a frequency of $10$ kHz, the total time of a single experiment is $n/(2f_{AC})\approx 2.4$ ms, shorter than the coherence time at 77 K but longer than the coherence time at room temperature. At room temperature, the spin state decoheres much faster than the experiment time, resulting in a factor of $\sim 3$ smaller fluorescence slope and corresponding magnetometric sensitivity than the 77 K case, in which the spin state does not decohere during the whole experiment time. In conclusion, the process of cooling enhances magnetometric sensitivities only for low enough frequencies, for which the experiment time at room temperature approaches $T_1$.      
\paragraph{}
To summarize, we have shown that, due to the accumulation of pulse imperfections, XY-based DD protocols are more efficient than for AC magnetometry than the CPMG protocol.  Additional concatenation may enhance the resulting sensitivities for measuring high-frequency fields, for which many DD pulses are required and robustness against pulse imperfections is essential. Since concatenation does not require significant experimental modifications over standard XY sequences, it may be implemented immediately for improving the performance of magnetometry at high AC frequencies. Moreover, the process of cooling is only efficient for measuring low-frequency AC fields, for which typical experiment times at room temperature approach the longitudinal relaxation time $T_1$.  
\section*{Acknowledgements}
We thank Linh My Pham for the fruitful discussions. This work has been supported in part by the EU CIG, the Minerva ARCHES award, the Israel Science Foundation (grant No. 750/14), the Ministry of Science and Technology, Israel, the CAMBR fellowship for Nanoscience and Nanotechnology, the AFOSR/DARPA QuASAR program, the German Federal Ministry of Education and Research (BMBF) within the Quantumtechnologien program (FKZ 13N14439), and  DFG through the DIP program (FO 703/2-1).

\bibliography{nvbibliography}

\end{document}